\documentclass[letterpaper,10pt]{article} 

\usepackage{osameet3} 

\usepackage{amsmath,amssymb}
\usepackage[colorlinks=true,bookmarks=false,citecolor=blue,urlcolor=blue]{hyperref} 

\begin{document}

\title{Efficient quasi-phase-matched frequency conversion in a lithium niobate racetrack microresonator}

\author{Jia-Yang Chen,  Zhao-hui Ma, Yong Meng Sua, and Yu-Ping Huang}
\address{Department of Physics, Stevens Institute of Technology, \\
Center for Quantum Science and Engineering, \\
Stevens Institute of Technology, 1 Castle Point Terrace, Hoboken, NJ 07030, USA}
\email{yuping.huang@stevens.edu}

\copyrightyear{2019}

\begin{abstract}
We demonstrate efficient second harmonic generation in a quasi-phase-matched, high quality factor ($Q_0 \approx 5.3\times 10^5$) racetrack microresonator. The observed normalized conversion efficiency is about $3.8\%~mW^{-1}$.
\end{abstract}


\section{Introduction}

 High quality factor (high-Q) lithium niobate on insulator (LNOI) microresonators \cite{Zhang:17} allow us to enhance nonlinear optical processes among the interacting lightwaves inside the cavity. High-order modal phase matching \cite{PhysRevApplied.11.034026} and cyclic phase matching \cite{cylic} have been utilized to successfully demonstrate efficient second harmonic generation (SHG). However, the former's performance is limited by its relatively low Q ($\sim 10^5$, due to the required narrow waveguide width $\sim 700 nm$ for achieving phase matching) and the poor mode overlapping ($<10\%$). While the latter is limited by the highly restricted mode overlapping between fundamental (pump) and fifth-order (second harmonic) modes. Here, we demonstrate for the first time, efficient second harmonic generation in a quasi-phase-matched high-Q ($Q_0 \approx 5.3\times 10^5$) racetrack microresonator with fundamental-modes matching. The recorded normalized conversion efficiency is $3.8\%~mW^{-1}$, which is more than 40 times higher than recently reported quasi-phase-matched lithium niobate microring\cite{Wolf:18}.
\section{Design, Fabrication and Measurement}
\begin{figure}[htbp]
  \centering
\includegraphics[width=6in]{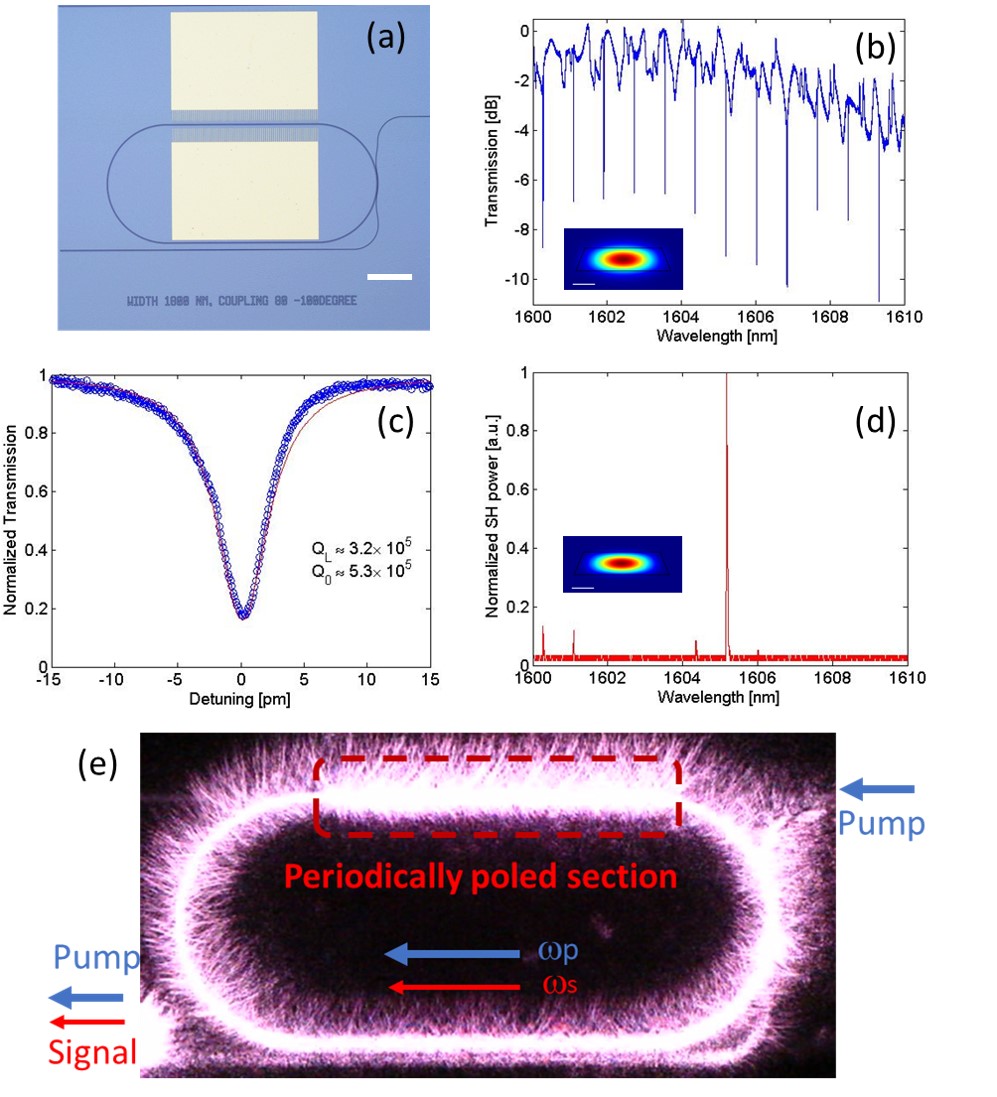}
\caption{\label{fig:figure1}LNOI racetrack microresonator design and characterization. (a) Microscope image of a patterned device. Waveguide top-width is 1.8 $\mu m$. The length of the straight waveguide is 300 $\mu m$ and the radius of the half-ring is 120 $\mu m$. Yellow region is the poling pattern for one straight waveguide within the racetrack microresonator. The scale bar is 100 $\mu m$. (b) The measured infrared spectrum for the racetrack microresonator. (c) Zoom-in spectrum of a resonance around 1605 nm. (d) SHG spectrum by sweeping the infrared laser from 1600 to 1610 nm. The insets in (b) and (d) are the simulated quasi-traverse-electric (TE$_{00}$) mode profiles for 1600 nm and 800 nm, respectively. The scale bar is 500 nm. (e) Microscope image of SHG in the racetrack microresonator with 100 $\mu W$ on-chip pump power. The dashed-line region is the periodically poled straight waveguide. More scattered SH light is observed due to higher loss in this section.}    
\label{figure1}
\end{figure}

To realize quasi-phase matching between fundamental modes along with the cavity enhancement, we design a periodically poled lithium niobate (PPLN) nanowaveguide within a racetrack microresonator(see Fig.~\ref{figure1}). Here for X-cut orientation, we utilize quasi-traverse-electric (TE$_{00}$) modes interaction to access the largest nonlinear coefficient $d_{33}$ ($\sim27 pm/V$) and their mode profiles are shown in the insets in Fig.~\ref{figure1}(b) and (d). To satisfy the phase matching condition in this racetrack microresonator, we not only need to fulfill quasi-phase matching for the PPLN section but also to achieve doubly-resonance for both pump and signal cavity modes. Due to different thermal-optic coefficients ($d \lambda /d T$) for those two modes, we can vary the global temperature of the chip to compensate their resonance mismatch.

The device is fabricated on a X-cut magnesium-doped LNOI wafer (NANOLN Inc.), which is a 500-nm thick LN thin film bonded on 2-$\mu m$ thermally grown silicon dioxide layer above a silicon substrate. Magnesium doped thin film is used to mitigate photorefractive effect induced by intense optical field of highly confined pump power. First, we use the similar process (as described in \cite{PPLN}) to create the poling region as shown in Fig.~\ref{figure1}(a). Then, additional EBL process is carried out to define the racetrack microresonator structure. The residual metallic poling pattern will be removed altogether during the following resist development and metal etch process. Then we obtain the high-Q racetrack microresonators ($Q_0 \approx 5.3\times 10^5$, see  Fig.~\ref{figure1} (c)) by carrying out a standard ion-milling process\cite{zenojyc,Chen:18}, with extracted average propagation loss to be $~0.7 dB/cm$.

\begin{figure}[htbp]
  \centering
\includegraphics[width=6in]{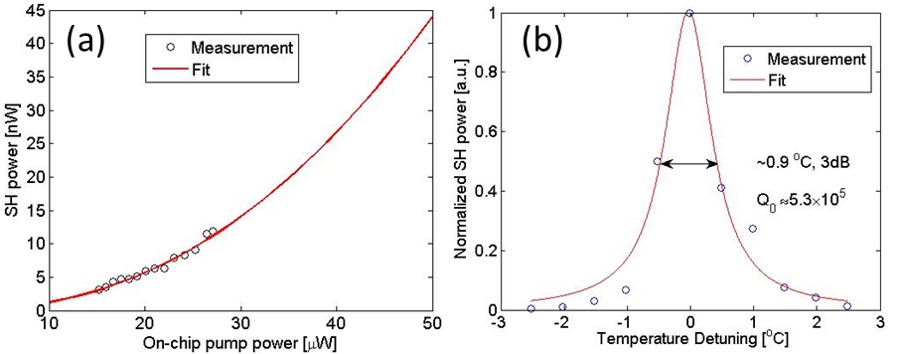}
\caption{\label{fig:figure2} (a) and (b) are the pump power and temperature dependency of SH power of the optical mode around 1605 nm with $Q_0 \approx 5.3\times10^5$. Here quadratic and Lorentzian fit is applied, respectively.}
\label{figure2}
\end{figure}

 To characterize the SHG efficiency, we use continuous-wave tunable laser (Santec 550, 1500-1630 nm) as pump laser along with a fiber polarization controller to excite the fundamental quasi-TE mode in the racetrack microresonator, as depicted in Fig.~\ref{figure1} (e). Two tapered fibers (2 $\mu$m spot diameter, OZ optics) serve as the input and output coupling with losses of 5 dB per facet at 1600 nm and 6 dB per facet at 800 nm, respectively.
 After sweeping the infrared laser and scanning the temperature, we identify an optimal cavity mode (i.e. 1605.199 nm at 34.2 $^\circ C$). For 1.1 mW pump power in the input fiber, we collect 700 nW SH power in the output fiber via visible power meter. Considering the coupling loss and near critical coupling for infrared cavity mode ($\sim$ 7dB deep), the estimated coupled-in pump power and on-chip SH power are 270 $\mu W$ and 2.79 $\mu$W, respectively. This corresponds to normalized SHG efficiency of $3.8\%~mW^{-1}$. To avoid thermal effect in this high-Q microresonator, we scan the pump power within few tens of micro-watts and collect the generated SH power. It shows nearly quadratic response ($\sim slope$ 2.2) as shown in Fig.~\ref{figure2}(a). Later on, we study the temperature dependency of the second harmonic generation for the optimized cavity mode (around 1605 nm). As shown in Fig.~\ref{figure2}(b), its 3 dB bandwidth is less than $1 ^\circ C$, which indicates the stringent requirement for meeting doubly-resonant condition with high-Q resonances. Nonetheless, adding a micro-heater on top of the microresonator will allow us to tune and lock doubly-resonant of fundamental and second harmonic modes the near future.  

\section{Conclusion}
In summary, we have realized efficient second harmonic generation in an optimally mode-matched, high-Q racetrack microresonator that is partially poled on chip. By utilizing cavity enhancement of high-Q racetrack microresonator, we achieved SHG efficiency as high as $3.8\%~mW^{-1}$. Our novel implementation of PPLN within a racetrack microresonator could enable a multitude of applications, such as quantum light generation and optical parametric amplification in integrated photonics circuits.

{\bf Acknowledgement:} The research was supported in part by National Science Foundation (Award \# 1842680).

\end{document}